\documentclass[aps,prl,preprint,groupedaddress]{revtex4-1}

\usepackage[dvips]{graphicx} 
\usepackage{amsfonts}
\usepackage{amssymb}
\usepackage{amscd}
\usepackage{amsmath}    
\usepackage{enumerate}
\usepackage{epsfig}
\usepackage{subfigure}
\usepackage{xcolor}
\usepackage{amsthm}
\usepackage{braket}

\begin{document}


\title{A resource-saving realization of the polarization-independent orbital-angular-momentum-preserving tunable beam splitter}



\author{Ya-Ping Li}
\author{Fang-Xiang Wang}
\author{Wei Chen}
\email{weich@ustc.edu.cn}
\author{Guo-Wei Zhang}
\author{Zhen-Qiang Yin}
\author{De-Yong He}
\author{Shuang Wang}
\author{Guang-Can Guo}
\author{Zheng-Fu Han}
\email{zfhan@ustc.edu.cn}
\affiliation{CAS Key Lab of Quantum Information, University of Science and Technology of China, Hefei 230026, People' s Republic of China}
\affiliation{State Key Laboratory of Cryptology, P. O. Box 5159, Beijing 100878, People' s Republic of China}

\date{\today}

\begin{abstract}
Tunable beam splitter (TBS) is a fundamental component which has been widely used in optical experiments. We realize a polarization-independent orbital-angular-momentum-preserving TBS based on the combination of modified polarization beam splitters and half-wave plates. Greater than 30 dB of the extinction ratio of tunableness, lower than $6\%$ of polarization dependence and more than 20 dB of the extinction ratio of OAM preservation show the relatively good performance of the TBS. In addition, the TBS can save about 3/4 of the optical elements compared with the existing scheme to
 implement the same function\cite{yang2016experimental}, which makes it have great advantages in scalable applications. Using this TBS, we experimentally built a Sagnac interferometer with the mean visibility of more than $99\%$, which demonstrates its potential applications in quantum information process, such as quantum cryptography. 
\end{abstract}


\maketitle

Orbital angular momentum (OAM) has recently attracted a growing interest as a high-dimensional resource for quantum information. Beams of OAM-carrying photons have an azimuthal phase dependence in the form of $e^{il\phi}$, where topological charge $l$ can take any integer value\cite{franke2008advances}\cite{yao2011orbital}\cite{willner2015optical}. Due to its unique property, OAM light can be applied to many fields, such as quantum entanglement\cite{weihs2001entanglement}\cite{leach2010quantum}, quantum simulation\cite{cardano2015quantum}\cite{luo2015quantum} and quantum communication\cite{molina2004triggered}\cite{vallone2014free}. However, dedicated techniques are necessary for manipulating and transmitting OAM of photons. Up to now, researchers have designed many optical elements for the translation and manipulation of OAM light, such as the OAM fiber\cite{ramachandran2009generation}\cite{bozinovic2013terabit}and Q-plate\cite{marrucci2006optical}\cite{d2012deterministic}. Meanwhile, tailored optical devices which can achieve fundamental optical functions are necessary as well.

Among these devices, tunable beam splitter (TBS) is an essential element to compose complex optical structures\cite{higgins2007entanglement}\cite{ma2011quantum}. There are three major methods to implement TBS. The first common realization of TBS is using the combination of a polarization beam splitter (PBS) and a half-wave plate (HWP), which has high extinction ratio, while it is polarization-dependent\cite{marcikic2003long}. Another type of polarization-independent TBS employs Mach-Zehender interferometers (MZIs) with high-speed modulators\cite{ma2011high}. Although this method can realize high-speed modulation, it is sensitive to external environment disturbance in different light path, such as the vibrations and temperature variations. Moreover, it has a relative low extinction ratio in some specific high-precision applications\cite{ma2011high}. Recently, Yang et al. realizes a polarization-independent TBS using the MZI composed of beam displacers (BD) and HWPs, in which the TBS has a relatively high polarization independence and high interference visibility. But it is sensitive to the phase in different paths and it has a complicated construction\cite{yang2016experimental}.

Here, we propose a polarization-independent OAM-preserving TBS based on HWPs and modified PBSs. The relatively low polarization dependence combining high extinction ratio of the TBS can save about 3/4 of the number of optical elements compared with the work in \cite{yang2016experimental}. The realization of Sagnac interferometer with interference visibility of above $99\%$ based on the TBS demonstrates its relatively good performance.
 
\begin{figure}[hbt]
\centering
\includegraphics{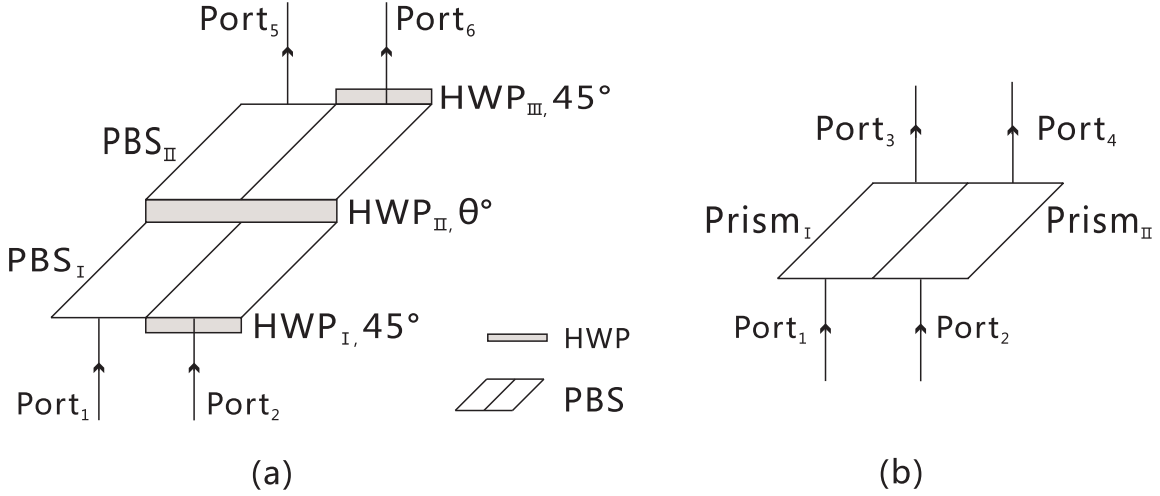}
\caption{Schematic diagrams of PBS and TBS. HWP, half wave plate; PBS, polarization beam splitter}
\label{fig:PBS_TBS}
\end{figure}

The structure of the TBS is shown in Fig. \ref{fig:PBS_TBS} (a). The TBS is composed of two modified PBSs and three HWPs, in which $HWP_{\uppercase\expandafter{\romannumeral2}}$ is used for adjusting the splitting ratio and $HWP_{\uppercase\expandafter{\romannumeral1}}$, $HWP_{\uppercase\expandafter{\romannumeral3}}$ are used to eliminate polarization dependence of the TBS. For a common cubic beam splitter (BS) or PBS, it changes the sign of topological charge $l$ of OAM light after a reflection. Therefore, the transmission matrix of BS or PBS is no longer a unitary matrix for OAM-carrying light, which may introduce inconvenience in some experiments\cite{mafu2013higher}\cite{zhang2016engineering}. The modified PBS used in the TBS is composed of two rhombic prisms (Sunlight Technology Co., H-K9L material), as shown in Fig. \ref{fig:PBS_TBS} (b). While a horizontal (vertical) polarization state enters into modified PBS from $Port_1$, it will exit from $Port_4$ ($Port_3$) after twice reflections and once transmission (only twice reflections). Therefore when photons that carry an OAM of $l\hbar$ enter into the PBS, the topological charge $l$ is preserved. Meanwhile the HWP does not change the topological charge. Thus, the TBS will be OAM-preserving. The property of polarization-independent of the TBS will be discussed in detail below. In a word, when an OAM photon with a quantum number of $l$ and arbitrary polarization state enters into the TBS from $Port_1$ or $Port_2$, it will exit from $Port_5$ and $Port_6$, preserving the original polarization state carrying OAM as shown in Fig. \ref{fig:PBS_TBS} (a).

The transmission matrix of the TBS can be specified using Dirac notation. When photons enter into $PBS_{\uppercase\expandafter{\romannumeral1}}$ from $Port_1$, and exit from $Port_3$ and $Port_4$, as shown in Fig. \ref{fig:PBS_TBS} (b). The operator of the $PBS_{\uppercase\expandafter{\romannumeral1}}$ can be described as:

\begin{equation}
M_{PBS_{\uppercase\expandafter{\romannumeral1}}} =  \Ket{h\otimes l,4}\Bra{h\otimes l,1}+\Ket{v\otimes l,3}\Bra{v\otimes l,1}   
\end{equation}

where 1 (3 and 4) means the $Port_1$ ($Port_3$ and $Port_4$) of the PBS. $h\otimes l$ ($v\otimes l$) denotes the incident horizontal (vertical) polarization state with OAM of $l\hbar$. 
 
Then the light enters into $HWP_{\uppercase\expandafter{\romannumeral2}}$ from $Port_3$ ($Port_4$) of $PBS_{\uppercase\expandafter{\romannumeral1}}$, whose operator can be described as:
\begin{align}
M_{HWP_{\uppercase\expandafter{\romannumeral2}}-3} &= cos2\theta\Ket{h\otimes l,3}\Bra{h\otimes l,3}+sin2\theta\Ket{h\otimes l,3}\Bra{v\otimes l,3} \\ 
              &+sin2\theta\Ket{v\otimes l,3}\Bra{h\otimes l,3}-cos2\theta\Ket{v\otimes l,3}\Bra{v\otimes l,3}   
\end{align}

\begin{align}
M_{HWP_{\uppercase\expandafter{\romannumeral2}}-4} &= cos2\theta\Ket{h\otimes l,4}\Bra{h\otimes l,4}+sin2\theta\Ket{h\otimes l,4}\Bra{v\otimes l,4} \\ 
              &+sin2\theta\Ket{v\otimes l,4}\Bra{h\otimes l,4}-cos2\theta\Ket{v\otimes l,4}\Bra{v\otimes l,4}   
\end{align}
where $\theta$ is the angle between the fast axis of HWP and the horizontal axis. $HWP_{\uppercase\expandafter{\romannumeral2}-3}$ ($HWP_{\uppercase\expandafter{\romannumeral2}-4}$) means the light enters into $HWP_{\uppercase\expandafter{\romannumeral2}}$ from $Port_3$ ($Port_4$).

When the light enters into the $PBS_{\uppercase\expandafter{\romannumeral2}}$ from $Port_3$ and $Port_4$, the operator of $PBS_{\uppercase\expandafter{\romannumeral2}}$ can be described as:
\begin{align}
M_{PBS_{\uppercase\expandafter{\romannumeral2}}} =  \Ket{h\otimes l,6}\Bra{h\otimes l,3}-\Ket{v\otimes l,5}\Bra{v\otimes l,3} 
           +\Ket{h\otimes l,5}\Bra{h\otimes l,4}+\Ket{v\otimes l,6}\Bra{v\otimes l,4}   
\end{align}
where the minus of the second item is due to the film coated on the left prism (here we assume the light reflected by coated film will introduce phase $\pi$). It is noted that the film is coated on the right prism in $PBS_{\uppercase\expandafter{\romannumeral1}}$. The operator of $HWP_{\uppercase\expandafter{\romannumeral3}}$ with $\theta$ equals to $45^\circ$ will be

\begin{equation}
M_{HWP_{\uppercase\expandafter{\romannumeral3}}} = \Ket{h\otimes l,6}\Bra{v\otimes l,6} + \Ket{v\otimes l,6}\Bra{h\otimes l,6}
\end{equation} 
Thus, we can deduce the operator of the TBS:

\begin{align}
M_{TBS}&=(1+M_{HWP_{\uppercase\expandafter{\romannumeral3}}})M_{PBS_{\uppercase\expandafter{\romannumeral2}}}(M_{HWP_{\uppercase\expandafter{\romannumeral2}-3}}+M_{HWP_{\uppercase\expandafter{\romannumeral2}-4}})M_{PBS_{\uppercase\expandafter{\romannumeral1}}} \\ 
	   &=cos2\theta\Ket{h\otimes l,5}\Bra{h\otimes l,1}+sin2\theta\Ket{h\otimes l,6}\Bra{h\otimes l,1}\\ 
	   &-cos2\theta\Ket{v\otimes l,5}\Bra{v\otimes l,1}+sin2\theta\Ket{v\otimes l,6}\Bra{v\otimes l,1}         
\end{align}
When an arbitrary polarization state carrying OAM of $l\hbar$ enters into the TBS from $Port_1$, the incident state can be described as:   
\begin{align}
\Ket{In}=\alpha\Ket{h\otimes l,1}+\beta\Ket{v\otimes l,1}
\end{align}
where $\alpha$ and $\beta$ are complex constants, and $\|{\alpha}\|^2+\|{\beta}\|^2=1.$
Thus, the output state should be:
\begin{align}
\Ket{Out} & =M_{TBS}\Ket{In} 																
          =cos2\theta(\alpha\Ket{h\otimes l,5}+\beta\Ket{v\otimes l,5})
          +sin2\theta(\alpha\Ket{h\otimes l,6}+\beta\Ket{v\otimes l,6})  
        \label{out}
\end{align}

According to Eq. (\ref{out}), the output states from $Port_5$ and $Port_6$ remain their original polarization and OAM states. The splitting ratio is determined by the $\theta$ of $HWP_{\uppercase\expandafter{\romannumeral2}}$. The same conclusion can be obtained by using a similar process when light enters into the TBS from $Port_2$.

\begin{figure}[htbp]
\centering
\includegraphics[width=\linewidth]{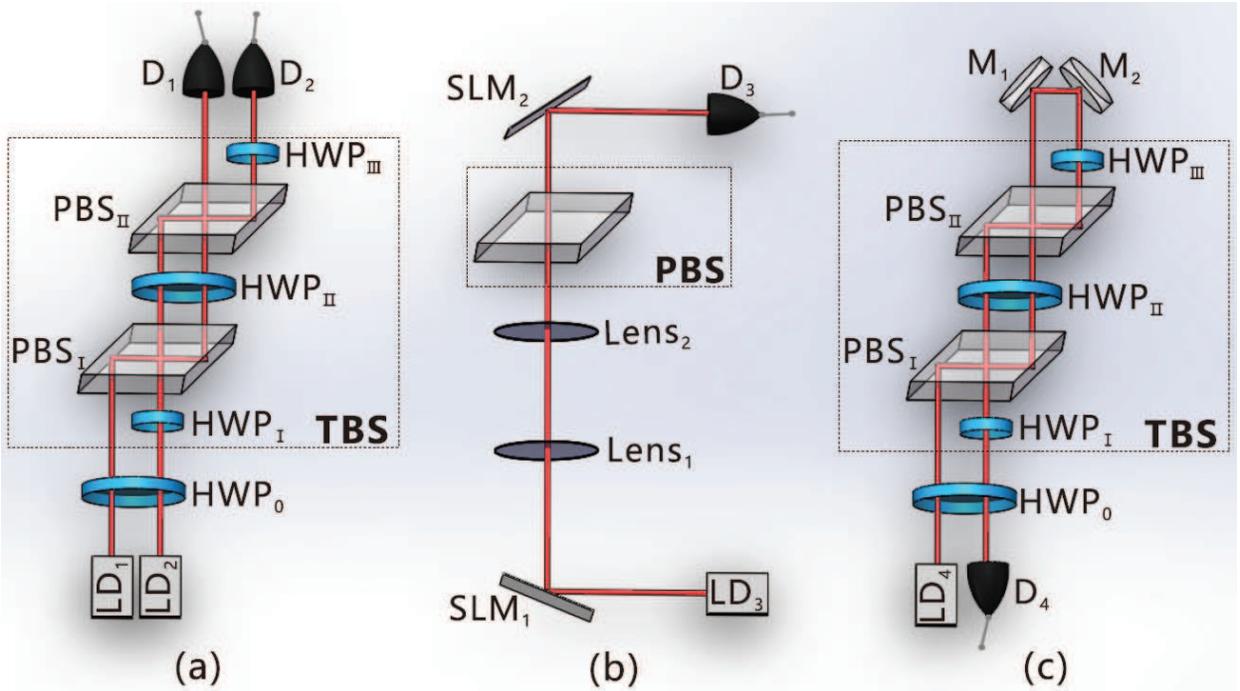}
\caption{(a) The experimental setup to measure the splitting ratio and the polarization independence. (b) 4F optical system for process tomography (figure not to scale). (c) The structure of the Sagnac interferometer. LD, laser diode; D, optical power meter; M, mirror; SLM, spatial light modulator; Lens, plano-vertex lens}\label{fig:Exp_Setup0321}
\end{figure}

\begin{figure}[htbp]
	\centering
	\includegraphics[width=\linewidth]{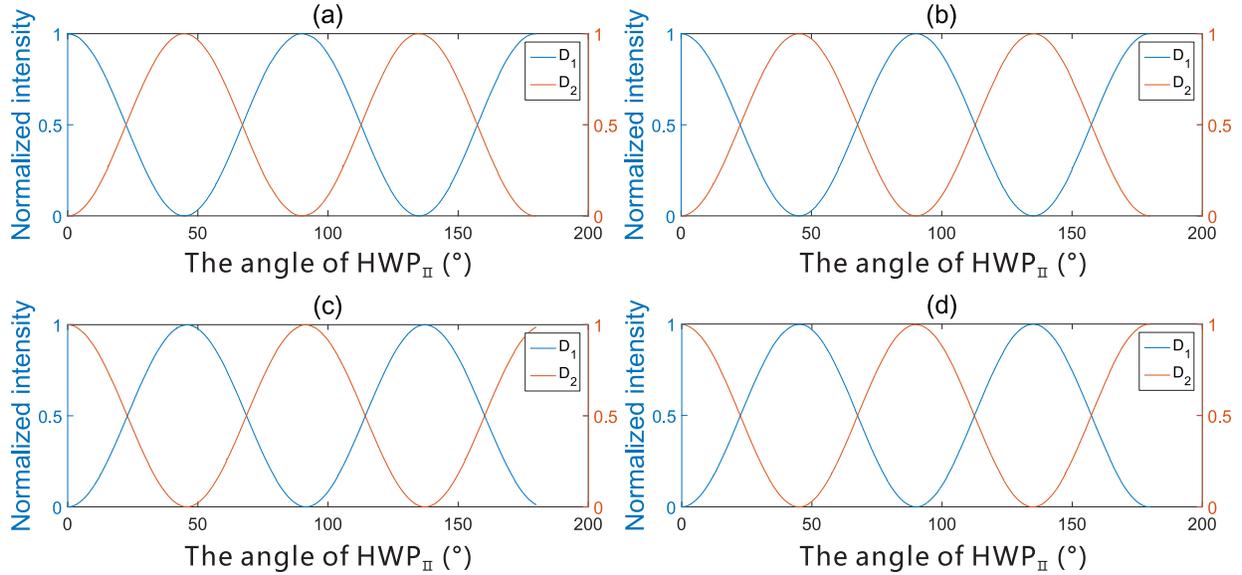}
	\caption{The tunable capacity of the TBS when the states (a) $\Ket{H}$ and (b) $\Ket{V}$ are prepared in $LD_1$, respectively. The states (c) $\Ket{H}$ and (d) $\Ket{V}$ are corresponding to the states prepared in $LD_2$.}
	\label{fig:Spli_Ratio}
\end{figure}

To verify the performance of the TBS, three experiments are conducted respectively as shown in Fig. \ref{fig:Exp_Setup0321}. The experimental setup to test the split-ratio tunableness of the TBS is shown in Fig. \ref{fig:Exp_Setup0321} (a). A horizontal polarized light beam emits from a continuous-wave laser with the wavelength of 780 nm enters into $HWP_{0}$ driven by a DC servo motor (PR50CC, Newport Co.), which can generate arbitrary linear polarization states. Here, we test the split-ratio tunableness when $\Ket{H}$ and $\Ket{V}$ enter into the TBS, respectively. For a fixed incident polarization state, we detect the light intensity by optical power meters $D_1$ and $D_2$ in the sample rate of 100 Hz when rotating the $HWP_{\uppercase\expandafter{\romannumeral2}}$ in the range of zero to $180^\circ$ with the precision of $0.1^\circ$. The intensities of the output light changed with the angle of $HWP_{\uppercase\expandafter{\romannumeral2}}$ are shown in Fig. \ref{fig:Spli_Ratio}. When the angle of $HWP_{\uppercase\expandafter{\romannumeral2}}$ is $0^\circ$ in Fig. \ref{fig:Spli_Ratio} (a), the intensity detected by $D_1$ and $D_2$ are maximum and minimum, respectively. It is a balanced beam splitter when the angle of $HWP_{\uppercase\expandafter{\romannumeral2}}$ is $22.5^\circ$, as shown in the intersection of the two curves. The criterion for evaluating the performance of the split-ratio tunableness of TBS is the extinction ratio (ER), which is defined as:

\begin{equation}
	ER=\frac{I_{max}}{I_{min}}  
\label{ER}
\end{equation}
where $I_{max}$ and $I_{min}$ refer to the maximum and minimum intensity of the output ports respectively. According to Eq. (\ref{ER}), the mean, maximum and minimum ERs of tunableness of the TBS are 34 dB, 30.7 dB and 40.1 dB, respectively. 

\begin{figure}[htbp]
	\centering
	\includegraphics[width=\linewidth]{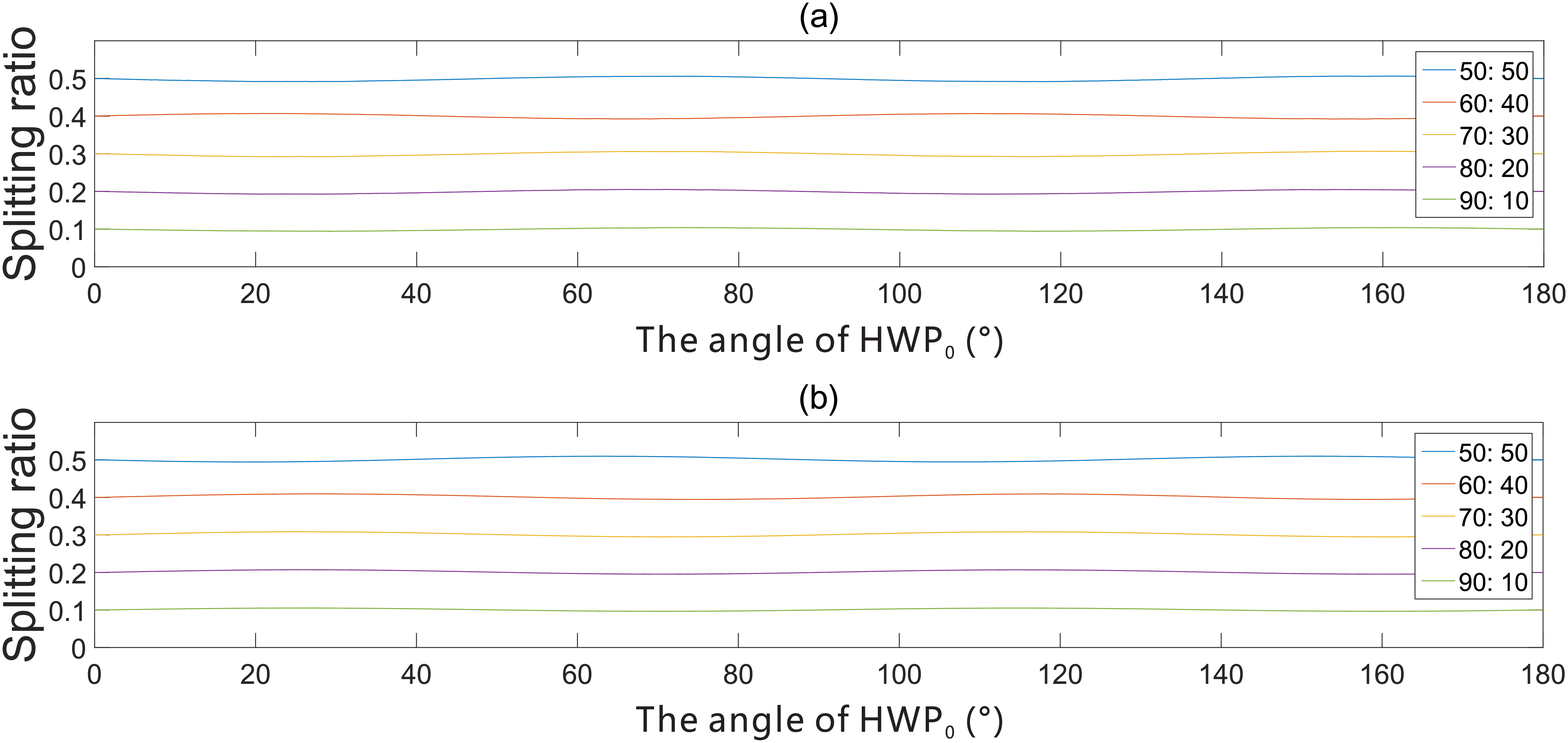}
	\caption{The polarization dependence of the TBS. The splitting ratios change with $HWP_0$ when light from (a) $LD_1$ and (b) $LD_2$.}
	\label{fig:DOPD}
\end{figure}

To verify the polarization independence of the TBS, we test the variation of the splitting ratio with the incident polarization states as shown in Fig. \ref{fig:Exp_Setup0321} (a). In the tests, various linear polarization states are prepared by rotating the $HWP_{0}$ in the range of zero to $180^\circ$ with the accuracy of $0.1^\circ$, and the light intensities are detected by optical power meters in the sample rate of 100 Hz. The results are shown in Fig. \ref{fig:DOPD}. To evaluate the performance of the polarization independence, we define the splitting ratio (SR) and polarization dependence (PD) as follows:

\begin{equation}
	SR=\frac{R}{R+T}  
\end{equation}

\begin{equation}
	PD= \frac{|SR_{exp}-SR_{th}|}{SR_{th}} 
	\label{DOPD}
\end{equation}

where R (T) denotes the intensity of reflected (transmission) light, and $SR_{exp}$ means the experimental splitting ratio with a maximum deviation from theoretical value, which is the worst case. $SR_{th}$ is the theoretical splitting ratio. In the experiment, we set the reflected light to be the weak light when TBS is in the unbalanced beam splitting. Therefore, the range of $SR_{th}$ will be 0.1 to 0.5 in our tests. The smaller the value of PD, the better the performance of polarization independence.

According to Eq. (\ref{DOPD}), the measurement results of the polarization dependence in different splitting ratios are shown in Tab. \ref{splitting ratio}. The worst case of the polarization dependence is lower than $6\%$ with a triple standard deviation of below $0.2\%$. With the decrease of SRs, PDs are gradually rising according to the experimental results. This is due to the more obvious influence of the intensity fluctuation to weak light. 

\begin{table}[htbp]
	\centering
	\caption{\bf Polarization dependence in different splliting ratios}
	\begin{tabular}{|c|c|c|c|c|c|} 
		\hline
	           $SR$            & $50: 50$      & $60: 40$     & $70: 30$	    & $80: 20$   & $90: 10$    \\
		\hline  
	           $PD_1$          & $1.73\%\pm 0.06\%$      & $2\%\pm 0.10\%$        & $2.63\%\pm 0.10\%$        & $3.68\%\pm 0.19\%$   & $5.75\%\pm 0.15\%$    \\
		\hline
        	   $PD_2$          & $2\%\pm 0.07\%$         & $2.42\%\pm 0.14\%$     & $2.88\%\pm 0.10\%$        & $3.69\%\pm 0.21\%$   & $5.35\%\pm 0.14\%$    \\
		\hline
	\end{tabular}
	\label{splitting ratio}
\end{table}

\begin{figure}[htbp]
\centering
\includegraphics[width=\linewidth]{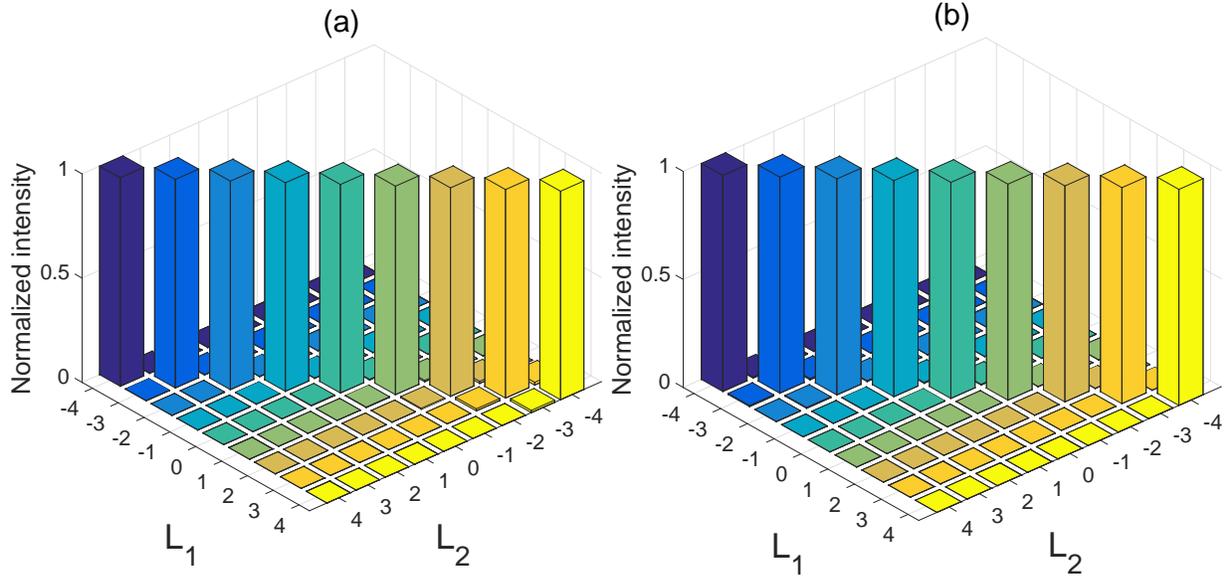}
\caption{The process tomography of (a) the PBS and (b) the TBS in the 4F optical system, respectively.}
\label{fig:OAM}. 
\end{figure}

To verify the OAM preservation performance of the TBS, a process tomography based on a 4F optical system has been proposed as shown in Fig. \ref{fig:Exp_Setup0321} (b). A 780 nm continuous-wave diode laser illuminates a spatial light modulator ($SLM_1$) to generate kinds of OAM light. A 4F optical system consisting of two plano-convex lenses with the focal length of 750 mm and a spatial filter with the aperture of 8 mm are employed to isolate the first order of the beam diffracted by the $SLM_1$. After transmitting through the PBS and TBS, OAM light is demodulated by $SLM_2$ and coupled into a single mode fiber (SMF) to be detected. If the forked hologram loaded by $SLM_2$ is identical to the $SLM_1$ ($l=l_1-l_2=0$), the OAM light will totally be converted to Gaussian beam theoretically\cite{mair2001entanglement}. However, imperfect devices will lead to the crosstalk from other OAM modes which can be seen as a criterion to evaluate the performance of OAM preservation of optical elements. In the process tomography, we prepare the OAM light with $l=0$, $\pm1$, $\pm2$, $\pm3$, $\pm4$ respectively, and then measure it with the same sets. The experimental results of the PBS and TBS are shown in Fig. \ref{fig:OAM}. In order to assess the OAM preservation performance, we define the ER of OAM preservation as follows:
\begin{equation}
	ER_{OAM}=\frac{I_i}{\sum_{k \ne i} I_k}  
\label{EXT_OAM}
\end{equation}
where $I_i$ and $I_k$ refer to the detected light intensity when the hologram loaded by $SLM_2$ is the identical order and other orders of OAM light, respectively. According to Eq. (\ref{EXT_OAM}), the ERs of more than 20 dB are obtained, which are close to the situations without PBS or TBS. These results demonstrate the PBS and TBS can preserve the OAM light well.

\begin{figure}[htbp]
	\centering
	\includegraphics[width=\linewidth]{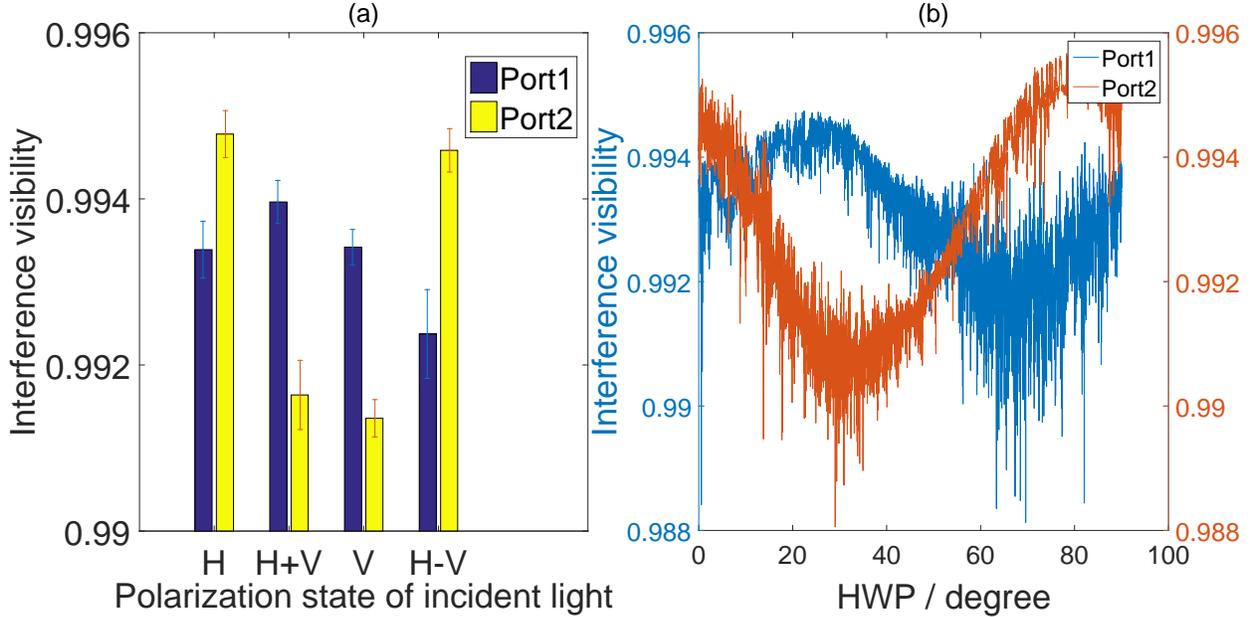}
	\caption{(a) The test of stability of the Sagnac interferometer. (b) The test of polarization independence of Sagnac interferometer.}
	\label{fig:Interference5050}
\end{figure}

Interferometer is a kind of important element in optical systems. Therefore, we built a Sagnac interferometer using this TBS to evaluate its performance in practical applications, as shown in Fig. \ref{fig:Exp_Setup0321} (c). In the Sagnac interferometer, the light emitted from $Port_{5}$ ($Port_{6}$) will be reflected by the mirrors $M_1$ and $M_2$ and returns to the TBS from $Port_{6}$ ($Port_{5}$), in which the two components throughing the same path form a closed loop. We test the visibility of the Sagnac interferometer during 30 minutes with the different incident polarization states. The interference visibility is defined as:
\begin{equation}
	V=\frac{I_{max}-I_{min}}{I_{max}+I_{min}}  
\end{equation} 
where $I_{max}$ and $I_{min}$ are the maximum intensity and minimum intensity of the output ports, respectively. As shown in Fig. \ref{fig:Interference5050} (a), the interference visibility of above $99\%$ with the triple standard deviations of below $0.2\%$ is obtained, which prove its good stability when the light enters into $Port_1$ and $Port_2$, individually. Then we test the interference visibility of different incident polarization state by rotating $HWP_0$ in a period from $0^\circ$ to $90^\circ$ with a step of $0.1^\circ$. The mean interference visibility for two input ports are all greater than $99\%$ as shown in Fig. \ref{fig:Interference5050} (b). The visibility of the incident polarization states $\Ket{H}$ and $\Ket{V}$ are closed, while the interference visibility reaches its maximum and minimum value when the incident polarization states are $\Ket{H+V}$ and $\Ket{H-V}$, respectively, as whown in Fig. \ref{fig:Interference5050} (b). This is due to the insertion loss of different polarization states in different ports and extinction ratio of PBS. The other curve of incident light from $Port_2$ is reflexive associative with the curve from $Port_1$, which is reasonable since the state $\Ket{H+V}$ from $Port_1$ is corresponding to the state $\Ket{H-V}$ from $Port_2$.

In conclusion, we realize a polarization-independent OAM-preserving TBS which needs much less optical devices comparing with the existing schemes. We experimentally evaluated the key parameters of the scheme, and demonstrated that the ERs of tunableness are greater than 30 dB, polarization dependence are lower than $6\%$ and ERs of OAM preservation are more than 20 dB. Using this TBS, we built a Sagnac interferometer with the mean visibility of above $99\%$, which makes it have potential to be utilized in kinds of quantum information processing. This resource-saving structure has the potential advantage to simplify the optical systems and be applied to the scalable applications\cite{wang2016scalable}\cite{krenn2017entanglement}. It is noted that the scheme can be implemented more compactly with emerging techniques such as integrated optics.   

This work has been supported by the National Natural Science Foundation of China (Grant Nos. 61675189, 61627820, 61622506, 61475148, 61575183), the National Key Research And Development Program of China (Grant Nos.2016YFA0302600, 2016YFA0301702), the "Strategic Priority Research Program(B)" of the Chinese Academy of Sciences (Grant No. XDB01030100).

\end{document}